\documentclass[%
 reprint,
superscriptaddress,
 amsmath,amssymb,
 aps,
]{revtex4-2}

\usepackage{amssymb,amstext,amsmath}
\usepackage{graphicx}
\usepackage{dcolumn}
\usepackage{bm} 
\usepackage{bbm} 
\usepackage{multirow}
\usepackage{bigstrut}
\usepackage{makecell,rotating}
\usepackage{mathrsfs}
\usepackage{booktabs}
\usepackage{threeparttable}
\usepackage{subfigure}
\usepackage{chngpage}
\usepackage{float}
\usepackage{color}
\usepackage{url}
\begin{document}

\preprint{APS/123-QED}

\title{Impact of misinformation in the evolution of collective cooperation}

\author{Yao Meng}
\affiliation{%
	Center for Systems and Control, College of Engineering, Peking University, Beijing 100871, China}

\author{Mark Broom}
\affiliation{%
	Department of Mathematics, City, University of London, Northampton Square, London EC1V 0HB, UK}

\author{Aming Li}
\thanks{amingli@pku.edu.cn}%
\affiliation{%
	Center for Systems and Control, College of Engineering, Peking University, Beijing 100871, China}
\affiliation{
	Center for Multi-Agent Research, Institute for Artificial Intelligence, Peking University, Beijing 100871, China}

\date{\today}

\begin{abstract}
	Human societies are organized and developed through collective cooperative behaviors, in which interactions between individuals are governed by the underlying social connections.
	It is well known that, based on the information in their environment, individuals can form collective cooperation by strategically imitating superior behaviors and changing unfavorable surroundings in self-organizing ways.
	However, facing the tough situation that some humans and social bots keep spreading misinformation, we still lack the systematic investigation on the impact of such proliferation of misinformation on the evolution of social cooperation.
	Here we study this problem by virtue of classical evolutionary game theory.
	We find that misinformation generally impedes the emergence of collective cooperation compared to scenarios with completely true information, although the level of cooperation is slightly higher when the benefits provided by cooperators are reduced below a proven threshold.
	We further show that this possible advantage shrinks as social connections become denser, suggesting that misinformation is more detrimental to the formation of collective cooperation when `social viscosity' is low.
	Our results uncover the quantitative effect of misinformation on the social cooperative behavior in the complex networked society, and pave the way for designing possible interventions to improve collective cooperation.
\end{abstract}

\maketitle
\section{Introduction}
The proliferation of misinformation on social networks---information that is incorrect or misleading---has become an escalating concern in public debate and academic research in recent years \cite{Pennycook2021a,Lazer2018,Brashier2021,Loomba2021,Scheufele2019}.
In addition to studying the diffusion of misinformation and seeking potential interventions and factors that might remedy misinformation among individuals, researchers further reported that misinformation significantly alters the behavior of individuals' decision-making \cite{Lazer2018,Pennycook2021,Pennycook2021a,VanderLinden2022,West2021,Scheufele2019,Tandoc2019,Vosoughi2018,Holme2019,del2016spreading,Kopp2018,Pennycook2022}.
Specifically, investigations showing that misinformation significantly alters the decision-making behavior of individuals have gained even more attention.
For instance, a randomised controlled trial shows that misinformation around COVID-19 vaccines induced a decline in the intent of $6.2$ percentage points in the UK and $6.4$ percentage points in the USA among those who stated that they would definitely accept a vaccine \cite{Loomba2021}.
In $2013$, a false tweet claiming that Barack Obama was injured in an explosion sent financial markets into a tailspin, wiping out $130$ billion in stock value \cite{stock}.

Despite that experimental research and survey reveal specific cases where misinformation significantly effects individual decision-making, how misinformation alters the evolution of collective cooperative behavior remains widely unknown.
To uncover the underlying mechanisms of how cooperative behavior evolves and persists, researchers have turned to the evolutionary game theory---a powerful framework and canonical paradigm for explaining the evolution of collective cooperation \cite{axelrod1981evolution,hofbauer1998evolutionary,nowak2006evolutionary,levin2020collective,Sigmund2010,rapoport1965prisoner}---in which interactions between individuals are determined by spatial relationships or social networks \cite{santos2005scale,hauert2004spatial,ohtsuki2006simple,taylor2007evolution,Santos2008Social,allen2017evolutionary,li2016evolutionary,li2020evolution}.
Traditionally, individuals are willing to change unfavorable surroundings by detecting other possible locations with higher expected payoffs, where the information that individuals rely on to make decisions is assumed to be completely true \cite{Vainstein2001,Vainstein2007,Helbing2009,Sicardi2009,Chen2012,Meloni2009}.
In reality, however, due to the proliferation of misinformation, individuals may receive false information about the alternative locations.
Here we first investigate the effect of misinformation on the evolution of social cooperation, and uncover how misinformation alters the migration trajectories of individuals and generally impedes the emergence of cooperation under different population densities.

\begin{figure*}[t]
	\centering
	\includegraphics[width=0.8 \textwidth]{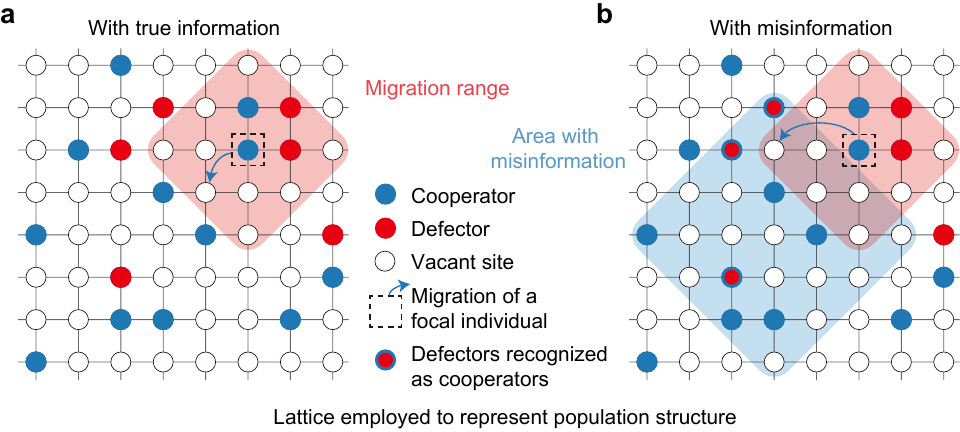}
	\caption{Illustration on the individual migration in areas with true information and misinformation.
		(a) Individuals are located on the lattice, where each site is either occupied by a cooperator (blue), or a defector (red) or remains vacant (white).
		In each round of game, individuals interact with their neighbours and accumulate payoffs.
		Then, individuals are allowed to migrate inside its
		migration range (red shaded area), which contains $n_M=13$ sites.
		To determine the location for migration, the individual (circled by dashed square) evaluates expected payoff of each empty site by fictitiously interacting with the site's neighbours.
		Afterwards, the individual chooses to move to the site with two cooperative neighbours (pointed by the solid arrow), which has the highest expected payoff among all available locations (see modelling framework).
		In the case that several sites have the same highest expected payoff, a random one is selected.
		And individuals will stay put if the highest expected payoff is lower than the individual's current payoff.
		(b) The area with misinformation is filled with blue.
		Defectors in this area are recognized as cooperators when individuals evaluate the fictitious expected payoffs in this area (circles with a red face and blue edge).
		Note that the site around with a cooperator and two defectors (pointed by the solid arrow) has the highest expected payoff and becomes the best choice in the individual's migration range.
	} \label{Fig1}
\end{figure*}

\section{Modeling framework}
Specifically, we employ the traditional square lattice with periodic boundary conditions containing $N=L\times L$ sites to represent the spatial social connections, where each site is either empty or occupied by a player.
The proportion of the number of individuals to all $N$ sites describes the density of population ($d$).
In a typical two-player prisoner's dilemma game, each individual could choose either cooperation (C) or defection (D).
A cooperator receives ``rewards" $R$ from mutual cooperation, while defectors obtain ``punishment" $P$ from mutual defection.
A defector attempting to exploit a cooperator obtains $T$ and leaves $S$ to its opponent cooperator.
Following the canonical practice \cite{Nowak92Nature,li2020evolution,Helbing2009}, a single parameter $T=b$ is employed here to depict the level of social dilemma for collective cooperation (namely $R=1, S=0, P=0.1$), where $b$ captures the advantages of defectors over cooperators ($1<b<2$).

We start by randomly placing the $Nd$ individuals on the lattice, where each individual adopts cooperation or defection equally.
At the beginning of each round of game, the individual at site $i$ interacts with their neighbours separately and accumulates the payoff $\pi_i(s_i)$, where $s_i$ indicates the strategy of the individual at site $i$.
Then a randomly selected individual $i$ on the lattice will decide whether to migrate to other empty sites based on the social information.
For different lattices, the migration range of each individual contains $n_M$ sites, consisting of the current site and the sites whose distance from this site are $1$ and larger (Fig.~\ref{Fig1}a).
The individual evaluates the fictitious expected payoffs $\tilde{\pi}_{j}(s_i)$ for any empty site $j$ in its migration range by hypothetically interacting with neighbours of the site $j$ using its current strategy $s_i$.
Then the individual will choose to migrate to $j$ with the highest expected payoff if $\tilde{\pi}_j(s_i)> \pi_i(s_i)$.
Afterward, the individual imitates the strategy of the best performing neighbour (namely, the neighbour having the highest payoff) if its own current payoff is lower.

Compared to the scenario where individuals obtain true information about the strategies of their neighbours around the alternative locations, misinformation appears within a certain fixed area.
In the area with misinformation, any defector is superficially recognized as a cooperator when individuals explore the expected payoffs of the corresponding empty sites (Fig.~\ref{Fig1}b).
Note that the real strategy of each individual can be correctly recognized when individuals imitate the strategies from their neighbours after migration.

\section{Results}
\begin{figure*}[t]
	\centering
	\includegraphics[width=0.8\textwidth]{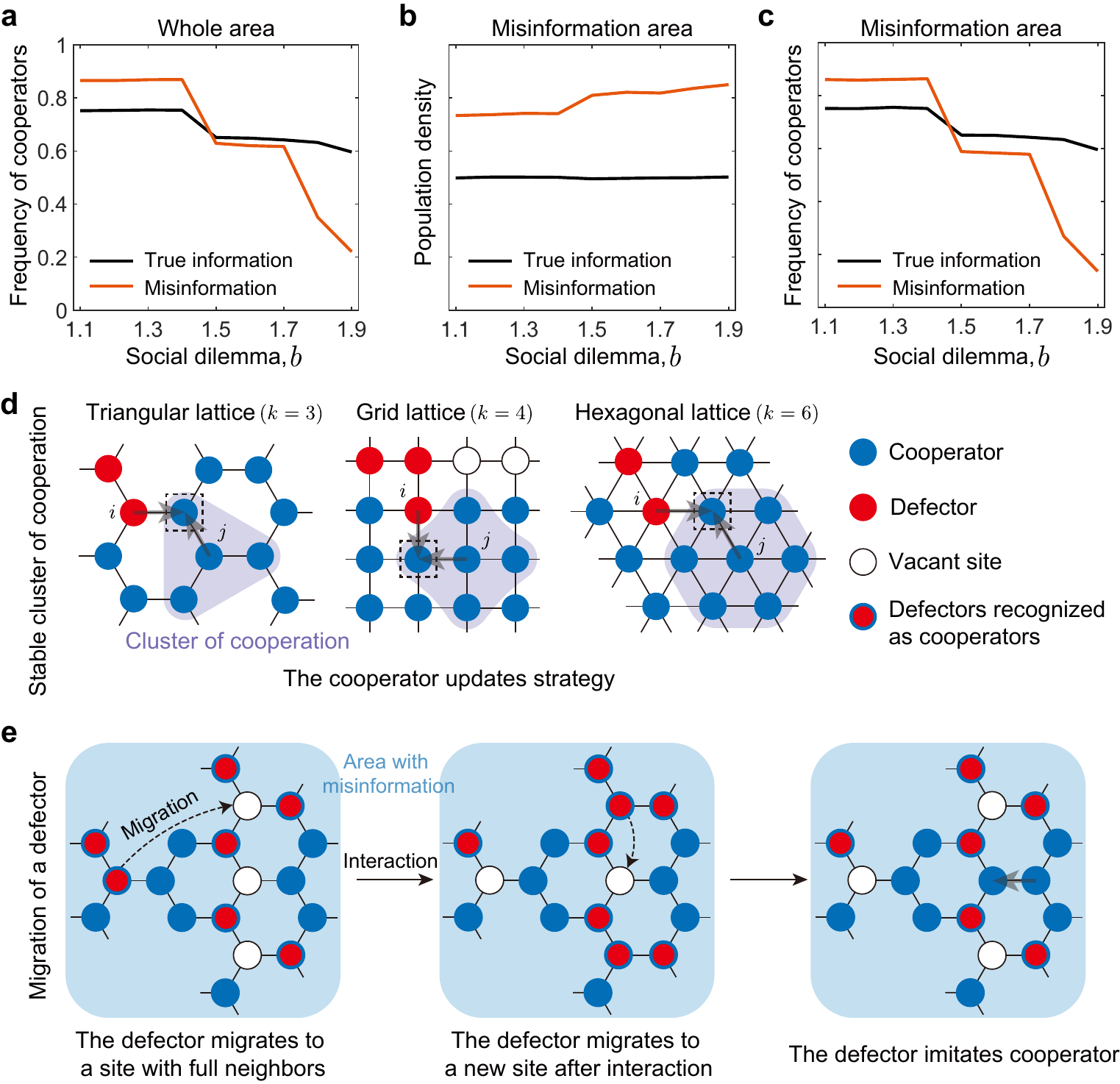}
	\caption{
		With misinformation, collective cooperation is promoted at weak social dilemma but inhibited at strong social dilemma.
		(a) For different levels of social dilemma ($b$), we plot the stable frequency of cooperators on triangular lattice. The frequency of cooperators with misinformation (orange) exceeds that with true information (black) when $b$ is small.
		As individuals are allowed to migrate during evolution, the population density in area with misinformation is higher than the that with true information at all range of $b$ as shown in (b).
		(c) The frequency of cooperators in the area with misinformation is also higher than that with true information in the same area at weak social dilemma, which is consistent with (a).
		The stable frequency of cooperators in (a)-(c) are obtained over $60$ independent runs, where the average frequency in each run is obtained over $2000$ rounds after a transient time of $2\times 10^5$ rounds.
		In the environment with misinformation,
		the configurations for forming stable cooperative clusters at weak dilemma for triangular, square and hexagonal lattices are illustrated in (d) where a defector could have $k-1$ cooperative neighbours with $k$ being the maximum number of neighbours.
		Taking the left panel as an example,
		the defector with two cooperative neighbours and the cooperator with full cooperative neighbours compete (indicated by gray arrow) for the site in the dashed box.
		To retain the cooperative cluster (purple shaded area), the payoff of the defector should be lower than the cooperator with full cooperative neighbours (Eq.~(1)).
		(e) Illustration of the process for a defector in the misinformation area to migrate and become a cooperator.
		The defector with two cooperative neighbours is misled to migrate to the site with full defective neighbours, which results in a reduction of the payoff after a round of game and further migration as shown in the middle panel.
		After moving to a site with a cooperative neighbour with the higher payoff, the defector imitates the cooperator in the right panel.
	}
	\label{Fig2}
\end{figure*}

We first explore how misinformation alters the evolution of social cooperation on lattices.
Intriguingly, the environment with misinformation even presents a slight enhancement of the frequency of cooperators at weak dilemma (small $b$) compared to the scenario with completely true information (Fig.~\ref{Fig2}a).
Since all players in the area with misinformation are recognized as cooperators when individuals evaluate the fictitious expected payoffs of the empty sites, they crowd into the misinformation area to pursue higher payoffs, and thus increase the local population density (namely, the density of individuals) (Fig.~\ref{Fig2}b).
Besides, we show that the level of cooperation in the misinformation area is also improved (Fig.~\ref{Fig2}c).

\begin{table*}[t]
	\centering
	\caption{Condition for maintaining stable cooperative clusters}
	\center
	\setlength{\tabcolsep}{6mm}
	\begin{threeparttable}
		\begin{tabular}{cccccc}
			\toprule
			$b$                        & $n_{\text{C}}^{\text{D}} =5$ & $n_{\text{C}}^{\text{D}}=4$ & $n_{\text{C}}^{\text{D}}=3$ & $n_{\text{C}}^{\text{D}}=2$ & $n_{\text{C}}^{\text{D}}=1$ \\ \midrule
			Triangular lattice ($k=3$) & --                           & --                          & --                          & $[1,1.45)$                  & $[1.45,2)$                  \\
			Square lattice ($k=4$)     & --                           & --                          & $[1,1.3)$                   & $[1.3,1.9)$                 & $[1.9,2)$                   \\
			Hexagonal lattice ($k=6$)  & $[1,1.18)$                   & $[1.18,1.45)$               & $[1.45,1.9)$                & $[1.9,2)$                   & --                          \\
			\bottomrule
		\end{tabular}
		In order to maintain the stable cooperative clusters, given the corresponding number of cooperative neighbours for the defector ($n_{\text{C}}^{\text{D}}$) at the boundary of the cooperative cluster, the range of the ``temptation" $b$ for different lattices is solved according to Eq.~(1).
	\end{threeparttable}
	\label{table_condition}
\end{table*}
Previous research uncover that the emergence of collective cooperation relies on the formation of stable cooperative clusters \cite{nowak2006evolutionary,hauert2004spatial,hofbauer1998evolutionary,santos2005scale,levin2020collective,ohtsuki2006simple}.
To investigate the mechanism by which misinformation affects the evolution of collective cooperation, we analyse whether stable cooperative clusters---in which the cooperators neither move nor change their strategy---can be formed during the evolutionary process with misinformation.
We define the migration-stable state to represent the state that individuals do not migrate after evaluating the expected payoffs of all empty sites among their migration range.
Mathematically, in this state,  for any individual at site $i$, the expected payoff $\tilde{\pi}_{j}\left(s_{i}\right)$ at any empty site $j$ within its migration range is no more than its current payoff $\pi_{i}(s_i)$, namely
\begin{equation}
	\tilde{\pi}_{j}\left(s_{i}\right) \leq \pi_{i}(s_i).\nonumber
\end{equation}
An individual can retain its strategy if its own payoff is higher than that of their all neighbours (otherwise it should have the same strategy as the neighbour who has the highest payoff), or the neighbour has a higher payoff possessing the same strategy as $i$.
Specifically, when the system reaches migration-stable state and all individuals satisfy one of the following conditions
\begin{equation}
	\begin{cases}s_{j}\in \{\text{C},\text{D}\}, & \text{if}~~\pi_{j \in \mathcal{N}_{i}}(s_j) \leq \pi_{i}(s_i)  \\ s_{j}=s_{i}, & \text{if}~~\pi_{j \in \mathcal{N}_{i}}(s_j) > \pi_{i}(s_i)\end{cases},
	\nonumber
\end{equation}
where $\mathcal{N}_{i}$ indicates the neighbour set of $i$.
In such case, the evolutionary process ends with a constant frequency of cooperators, and the stable clusters of cooperators are maintained (Supplementary Figs.~1 and~2).

Next, we provide the condition for maintaining stable clusters of cooperators for different lattices (Fig.~\ref{Fig2}d).
During the evolutionary process, to prevent defectors from invading cooperative clusters, the defector (site $i$ in Fig.~2d) on the boundary of cooperative cluster should have lower payoff than the cooperator (site $j$ in Fig.~2d) with full cooperative neighbours, which means
\begin{equation}
	n_{\text{C}}^\text{D} b+\left(k-n_{\text{C}}^\text{D}\right)   P<k R.
	\label{condition}
\end{equation}
Here $n_{\text{C}}^\text{D}$ represents the number of cooperative neighbours for the defector ($i$), and $k$ captures the number of neighbours for each site on the lattice.
The left side represents the payoff of a defector interacting with $n_{\text{C}}^\text{D}$ cooperators and $k-n_{\text{C}}^\text{D}$ defectors, and the right side represents the payoff of a cooperator interacting with $k$ cooperative neighbours.
To satisfy Eq.~(1), a defector who achieves the maximum payoff can have $k-1$ cooperative neighbours, which results in the threshold $b^*=(kR-P)/(k-1)$, below which the defector on the boundary of a stable cooperative cluster can have no more than $k-1$ cooperative neighbours (Table~1).
From the perspective of forming and maintaining stable cooperative clusters, we further uncover that when the ``temptation" $b$ is weak (i.e., weak dilemma of collective cooperation), namely lower than the threshold $b^*$, the level of cooperation is maintained in an environment with misinformation and is even slightly higher than that with completely true information (Fig.~2a).

\begin{figure*}[t]
	\centering
	\includegraphics[width=0.8 \textwidth]{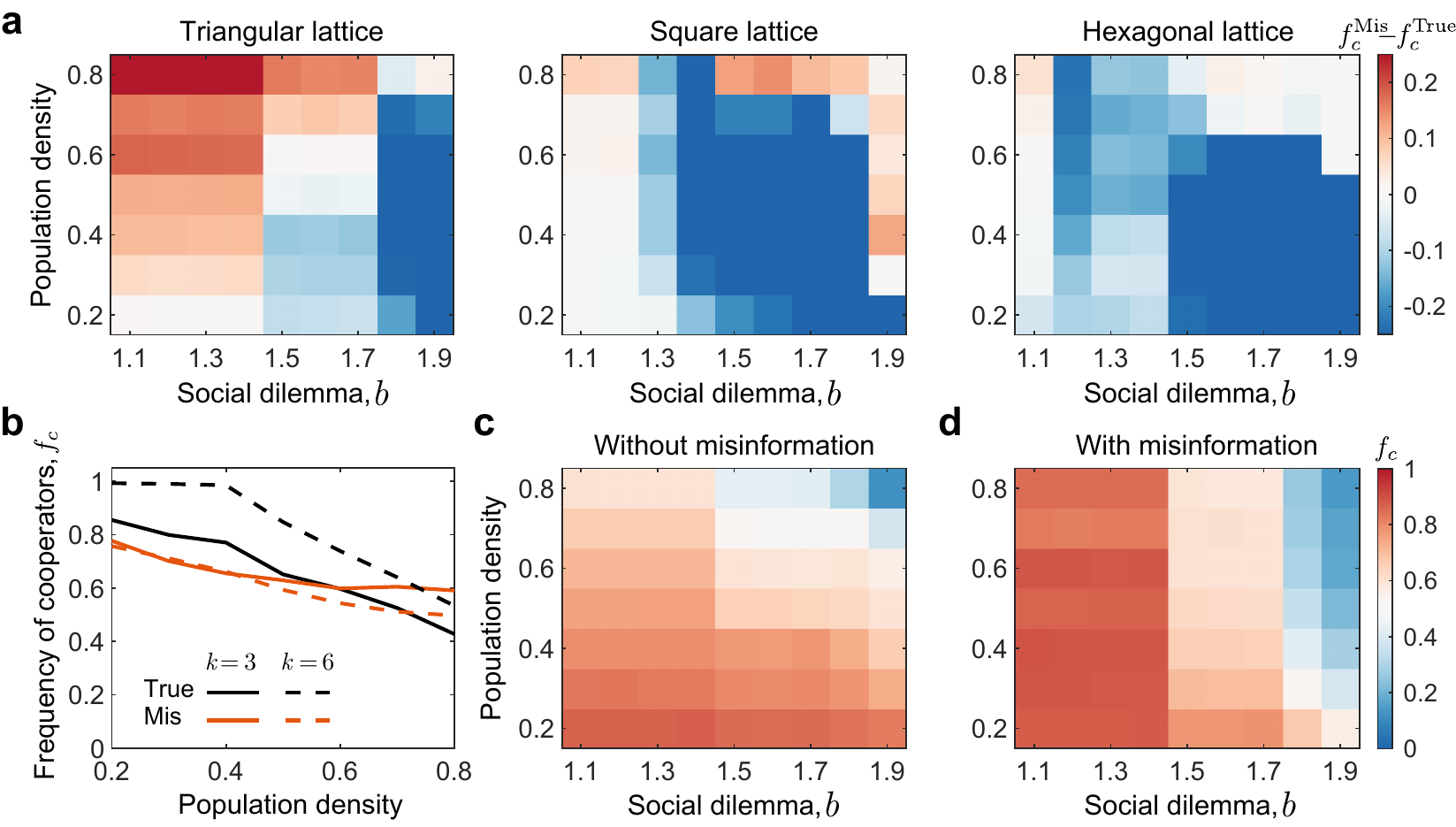}
	\caption{Effect of population density on collective cooperation with misinformation and true information.
	We compare the frequency of cooperators during the evolution with misinformation ($f_c^{\text{Mis}}$) and that with true information ($f_c^{\text{True}}$) on triangular ($k=3$), square ($k=4$), hexagonal ($k=6$) lattices with periodic boundaries with population density ranging from $0.2$ to $0.8$.
	The difference with individual migration range of $n_M=121$ is plotted in (a) where the red color indicates $f_c^{\text{Mis}}>f_c^{\text{True}}$ and blue indicates $f_c^{\text{Mis}}<f_c^{\text{True}}$.
	(b) We show the stable frequency of cooperators ($f_c$) through population density with true information (black) and misinformation (orange) on triangular (solid line) and hexagonal lattices (dashed line) at $b=1.5$.
	Furthermore, we plot $f_c$ for the evolutionary process with true information in (c) and that with misinformation in (d) over different population densities and levels of social dilemma, which indicates that the variations of population density generally leave weak effect on the evolution of cooperation with misinformation.
	The total number of nodes on triangular, square and hexagonal lattices are $2380, 2401, 2312$ respectively.
	}
	\label{Fig3}
\end{figure*}

We next reveal the mechanism by which collective cooperation in the environment with misinformation is maintained under weak social dilemma by analyzing the fate of a defector during the evolutionary process.
For the defector on the boundary of a stable cooperative cluster (who are allowed to have $k-1$ cooperative neighbours when $b<b^*$), the misinformation misleads it to site with full neighbours in the area with misinformation to pursue higher payoff (left panel in Fig.~\ref{Fig2}e).
However, this brings an even lower payoff for the defector in the subsequent interaction and drives it to continue its migration to a new site (middle panel in Fig.~\ref{Fig2}e).
Finally, the defector imitates the cooperative neighbour who has higher payoff, and a stable cooperative cluster is thus established (right panel in Fig.~\ref{Fig2}e).
For the whole population, when there are no empty sites with $k$ neighbours, individuals with $k-1$ cooperative neighbours will then stay put, leading the emergence of the migration-stable state.
At this point, the number of defectors has decreased since defectors with $k-1$ cooperative neighbours have been driven to migrate from original sites, and ultimately imitate their cooperative neighbours in new locations (Fig.~2e).
We numerically confirm the threshold $b^*$ for slightly promotion of collective cooperation with misinformation at a weak social dilemma on the triangular, square and hexagonal lattices with different population densities (Fig.~\ref{Fig3}a).
Despite that the level of collective cooperation is slightly enhanced when $b<b^*$, we next show that misinformation impedes the emergence of cooperation in strong dilemma of social cooperation ($b\geq b^*$), especially for dense social connections (Fig.~\ref{Fig3}a).
The unending migration induced by misinformation prevents the formation of stable cooperative clusters.
From Eq.~(\ref{condition}), we know that when $b\geq b^*$, the defectors on the boundary of a stable cooperative cluster should have no more than $k-2$ cooperative neighbours in order to sustain the stable cooperative cluster.
Due to the existence of the misinformation, individuals keep migrating to sites with $k-1$ or $k$ neighbours, which actually can not sustain the formation of stable cooperative clusters.
Therefore, the migration of individuals never stops due to the considerable amount of empty sites with $k-1$ neighbours, leaving no possibility to form stable cooperative clusters (Supplementary Figs.~3--5).
During the evolutionary game process with misinformation, mobile individuals will fill the entire misinformation area,
resulting in the general reduction of the level of cooperation for large $b$ (Fig.~\ref{Fig3}a).
As the social connections becomes denser (from the left to right panel in Fig.~3a), the threshold $b^*$ decreases (Eq.~(1)) and then the negative impact of misinformation is enlarged.

Furthermore, we show the impact of population density on the emergence of collective cooperation.
The high population density further shrinks the alternative empty sites for migration in the environment with true information.
With strong dilemmas, individuals tend to stay put and imitate their neighbours, which accelerates the disintegration of cooperative clusters (Supplementary Figs.~6--8).
This explains why the frequency of cooperation decreases drastically as population density increases in strong dilemmas (black lines in Fig.~\ref{Fig3}b, and Fig.~\ref{Fig3}c).
In contrast, in the environment with misinformation, the variation of population density causes a relatively mild change in the level of collective cooperation since the mobile individuals will fully fill the misinformation environment irrespective the population density when $b\geq b^*$ (red lines in Fig.~\ref{Fig3}b, and Fig.~\ref{Fig3}d).

\begin{figure*}[t]
	\centering
	\includegraphics[width= 0.8\textwidth]{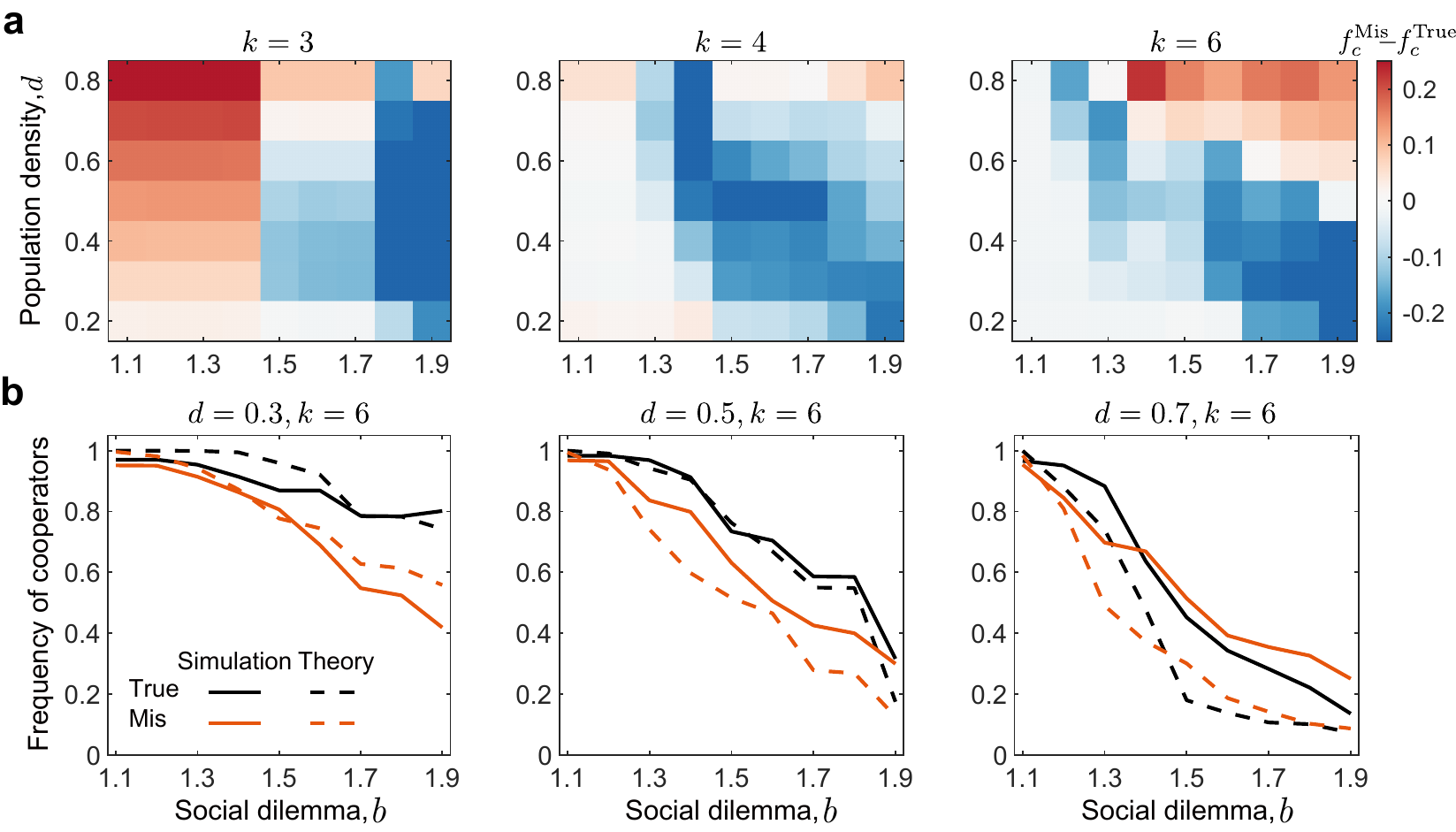}
	\caption{Our results hold true for large random regular social connections.
	(a) We show the heatmap of the difference between frequency of cooperators with misinformation ($f_c^{\text{Mis}}$) and true information ($f_c^{\text{True}}$) on random regular social connections with average number of neighbours of $k=3,4,6$ respectively.
	Consistent with the results we have shown for lattices, the emergence of cooperation is promoted with misinformation in weak dilemmas and sparse networks, while $f_c^{\text{Mis}}<f_c^{\text{True}}$ can be observed as $b$ increases and the networks become denser.
	(b) We obtain the theoretical stable frequency of cooperators (dashed line) through pair approximation on random regular social connections with $k=6$.
	The theoretical (dashed line) and numerical results (solid line) with true information (black) and misinformation (orange) at population density $d=0.3, 0.5, 0.7$ are shown from left to right respectively.
	The total number of nodes of the population represented by random regular network is $2400$.
	} \label{Fig4}
\end{figure*}

By numerical simulations and theoretical analysis, we check the robustness of our results for random regular networks (Fig.~\ref{Fig4}a, see Methods).
Figure~\ref{Fig4}b shows the theoretical approximation and simulation results at different population densities on random regular graphs, where we demonstrate that the emergence of social cooperation is generally impeded by misinformation both theoretically and numerically.
Compared to previous results for lattices, the level of cooperation in the scenario with misinformation is slightly improved on the random regular networks at high densities and strong dilemmas.
This is due to the shorter characteristic path length on random regular graphs compared with lattices, where a small migration range is enough for individuals to migrate out of the area with misinformation to form mobile cooperative clusters in a single step.

\section{Discussion}

Overall, misinformation misleads individuals in their decision and leads to ineffective migration, which ultimately inhibits the formation of stable clusters of collective cooperation in a wide range of social dilemma and population densities (Supplementary Figs.~9--14).
A promising direction for future research is the effects of misinformation on coevolutionary games where both the environment and strategies evolve over time \cite{perc2010coevolutionary,li2020evolution}.
Indeed, during the evolution of strategies, there are many factors in the environment that can feed back into the behavior of individuals \cite{pacheco2006coevolution,perc2010coevolutionary}.
Considering that previous studies have assumed that individuals receive completely true information from the environment, the coevolutionary games affected by misinformation may be a more realistic reflection of how individual behavior---in particular, collective social cooperation---evolves in practice.

Current efforts are mainly devoted to exploring effective interventions to mitigate the negative effects of misinformation on individuals as well as the society, by understanding what the misinformation is, how it operates, and designing the mechanism for suppressing the spread of misinformation \cite{Scheufele2019,Lazer2018,Brashier2021}.
However, the challenge remains to create a new system of safeguards to put an end to the impact of misinformation in collective social cooperation \cite{Lazer2018}.
Our study sheds new light on the effect of misinformation on the emergence of high level of collective cooperation.
We uncover that defectors on the boundary of the cooperative cluster can be misled to locations with lower real payoffs, providing the opportunity to maintain stable cooperative clusters by controlling the ``temptation" of social defection below a proven threshold.
Our findings suggest that even in environments with misinformation, effective intervention and organization by reducing the profit of defectors can be designed to dwindle the potential impact of misinformation in the emergence of collective social cooperation.


\section*{Appendixes}
\subsection{Pair approximation}
We represent the probability of a site occupied by cooperator ($\text{C}$), defector ($\text{D}$) or remained vacant ($\text{V}$) as $p_c$, $p_d$ and $p_v$ seperately, and denote the probability of finding a cooperator with a neighbouring cooperator by $p_{cc}$.
Similarly, the probability of the configuration of $\text{CD}$, $\text{CV}$, $\text{DD}$ pairs are defined as $p_{cd}, p_{cv}$ and $p_{dd}$, respectively.
Together with these four pair configuration probabilities, the evolutionary dynamics driven by migration with their expected change rate, which is calculated by considering all neighbouring configurations at original sites and migrated sites.
The trajectory that a cooperator at site $i$, with $n^o_c$ cooperative neighbours and $n^o_d$ defective neighbours, staying put and imitating its defective neighbour occurs at the probability
\begin{equation}
	\begin{aligned}
		\mathrm{P}_{\text{C} \rightarrow \text{D}}^{i \rightarrow i}\left(n_{c}^{o}, n_{d}^{o} ; n_{c}^{m}, n_{d}^{m}\right) & =
		p_{c} P_\text{C}\left(n_{c}^{o}, n_{d}^{o}\right) P_{M \max }^{(\text{C})}\left(n_{c}^{m}, n_{d}^{m}\right)                                                                                                                                    \\
		                                                                                                                     & \mathbbm{1}{\left( \tilde{\pi}_{\text{C}}\left(n_{c}^{m}, n_{d}^{m}\right) \leq \pi_{i}\right)} p_{\max }^{d},\nonumber
	\end{aligned}
\end{equation}
where $P_\text{C}\left(n_{c}^{o}, n_{d}^{o}\right)$ represents the probability of a cooperator having $n_c^o$ cooperators and $n_d^o$ defectors out of $k$ neighbours at its original site.
$\tilde{\pi}_{\text{C}}\left(n_{c}^{m}, n_{d}^{m}\right)$ captures the fictitious expected payoff of a cooperator interacting with $n_{c}^{m}$ cooperators and $n_{d}^{m}$ defectors.
$\mathbbm{1}(\cdot)$ is the indicator function.
$P_{M \max }^{(\text{C})}\left(n_{c}^{m}, n_{d}^{m}\right)$ indicates the probability of an empty site with $n_c^m$ cooperative and $n_d^m$ defective neighbours having the maximum fictitious payoff for a cooperator in its migration range, which is
\begin{equation}
	\begin{aligned}
		 & P_{M \max }^{(\text{C})}\left(n_{c}^{m}, n_{d}^{m}\right) =
		P_{\text{V}}\left(n_{c}^{m}, n_{d}^{m}\right)                  \\& \frac{\left(\left[\mathrm{P}\left(\tilde{\pi}_{j} \leq \tilde{\pi}_{\text{C}}\left(n_{c}^{m}, n_{d}^{m}\right)\right)\right]^{n_{\text{V}}} - \left[\mathrm{P}\left(\tilde{\pi}_{j} < \tilde{\pi}_{\text{C}}\left(n_{c}^{m}, n_{d}^{m}\right)\right)\right]^{n_{\text{V}}}\right) }{\sum_{n_{c}^{u}, n_{d}^{u}} P_{\text{V}}\left(n_{c}^{u}, n_{d}^{u}\right) \mathbbm{1}{\left(\tilde{\pi}_{\text{C}}\left(n_{c}^{u}, n_{d}^{u}\right)=\tilde{\pi}_{\text{C}}\left(n_{c}^{m}, n_{d}^{m}\right)\right)}},\nonumber
	\end{aligned}
\end{equation}
where $n_\text{V}$ is the number of empty sites in migration range.
Analogously, $P_\text{V}\left(n_{c}^{m}, n_{d}^{m}\right)$ represents the probability of a vacant site having $n_c^m$ cooperators and $n_d^m$ defectors out of $k$ neighbours.
We define $p_{\max }^{d}$ as the probability that the highest payoff is from a defector among the new neighbours, which is $$p_{\max }^{d}=\mathrm{P}\left(\pi_{l}<\pi_{h}, \pi_{i}<\pi_{h}, \forall l \in \mathcal{N}_{j}, l \neq h, s_{h}=\text{D} \right).$$
The probability of a cooperator at site $i$ (with $n^o_c$ cooperative neighbours and $n^o_d$ defective neighbours) migrating to an empty site $j$ (with $n^m_c$ cooperative neighbours and $n^m_d$ defective neighbours) and keeping cooperation is
\begin{equation}
	\begin{aligned}
		&\mathrm{P}_{\text{C} \rightarrow \text{V} \rightarrow \text{C}}^{i \rightarrow j}\left(n_{c}^{o}, n_{d}^{o} ; n_{c}^{m}, n_{d}^{m}\right)  =p_{c} P_{\text{C}}\left(n_{c}^{o}, n_{d}^{o}\right)  \\ & P_{M \max }^{(\text{C})}\left(n_{c}^{m}, n_{d}^{m}\right) \mathbbm{1}\left(\tilde{\pi}_{\text{C}}\left(n_{c}^{m}, n_{d}^{m}\right)>\pi_{i}\right)\left(p_{\max }^{o}+p_{\max }^{c}\right), \nonumber
	\end{aligned}
\end{equation}
where $p_{\max }^{o}=\mathrm{P}\left(\pi_{l} \leq \pi_{i}, \forall l \in \mathcal{N}_{j}\right)$ denotes the probability that individual has the highest payoff among its new neighbours,
and $$p_{\max }^{c}=\mathrm{P}\left(\pi_{l}<\pi_{h}, \pi_{i}<\pi_{h}, \forall l \in \mathcal{N}_{j}, l \neq h, s_{h}=\text{C}\right)$$
indicates the probability that the highest payoff is from a cooperator among the new neighbours.
Analogously, the probability of a cooperator at site $i$ (with $n^o_c$ cooperative neighbours and $n^o_d$ defective neighbours) migrating to empty site $j$ (with $n^m_c$ cooperative neighbours and $n^m_d$ defective neighbours) and imitating its new defective neighbour is
\begin{equation}
	\begin{aligned}
		\mathrm{P}_{\text{C} \rightarrow \text{V} \rightarrow \text{D}}^{i \rightarrow j}\left(n_{c}^{o}, n_{d}^{o} ; n_{c}^{m}, n_{d}^{m}\right) & =p_{c} P_{\text{C}}\left(n_{c}^{o}, n_{d}^{o}\right) P_{M \max }^{(\text{C})}\left(n_{c}^{m}, n_{d}^{m}\right)    \\
		                                                                                                                                          & \mathbbm{1}{\left(\tilde{\pi}_{\text{C}}\left(n_{c}^{m}, n_{d}^{m}\right)>\pi_{i}\right)} p_{\max }^{d}.\nonumber
	\end{aligned}
\end{equation}
The probability for a defector changing its neighbour configuration is similarly obtained as $\mathrm{P}_{\text{D} \rightarrow \text{C}}^{i \rightarrow i}$, $\mathrm{P}_{\text{D} \rightarrow \text{V} \rightarrow \text{C}}^{i \rightarrow j}$ and $\mathrm{P}_{\text{D} \rightarrow \text{V} \rightarrow \text{D}}^{i \rightarrow j}$.

\subsection{Stable frequency of cooperators}
The expected change rate of the pair configuration probability is given by
\begin{equation}
	\begin{aligned}
		\dot{p}_{c c}=
		 & \frac{2}{N k} \sum_{n_{c}^{o}, n_{d}^{o} ; n_{c}^{m}, n_{d}^{m}}(\mathrm{P}_{\text{C} \rightarrow \text{D}}^{i \rightarrow i} \Delta n_{c c}^{\text{C} \rightarrow \text{D}}+\mathrm{P}_{\text{D} \rightarrow \text{C}}^{i \rightarrow i} \Delta n_{c c}^{\text{D} \rightarrow \text{C}} \\ &+\mathrm{P}_{\text{C} \rightarrow \text{V} \rightarrow \text{D}}^{i \rightarrow j} \Delta n_{c c}^{\text{C} \rightarrow \text{V} \rightarrow \text{D}}+ \mathrm{P}_{\text{C} \rightarrow \text{V} \rightarrow \text{C}}^{i \rightarrow j} \Delta n_{c c}^{\text{C} \rightarrow \text{V} \rightarrow \text{C}} \\
		 & +\mathrm{P}_{\text{D} \rightarrow \text{V} \rightarrow \text{D}}^{i \rightarrow j} \Delta n_{c c}^{\text{D} \rightarrow \text{V} \rightarrow \text{D}}+\mathrm{P}_{\text{D} \rightarrow \text{V} \rightarrow \text{C}}^{i \rightarrow j} \Delta n_{c c}^{\text{D} \rightarrow \text{V} \rightarrow \text{C}}), \nonumber
	\end{aligned}
\end{equation}

\begin{equation}
	\begin{aligned}
		\dot{p}_{c d}=
		& \frac{1}{N k} \sum_{n_{c}^{o}, n_{d}^{o} ; n_{c}^{m}, n_{d}^{m}}(\mathrm{P}_{\text{C} \rightarrow \text{D}}^{i \rightarrow i} \Delta n_{c d}^{\text{C} \rightarrow \text{D}}+\mathrm{P}_{\text{D} \rightarrow \text{C}}^{i \rightarrow i} \Delta n_{c d}^{\text{D} \rightarrow \text{C}}   \\
		& +\mathrm{P}_{\text{C} \rightarrow \text{V} \rightarrow \text{D}}^{i \rightarrow j} \Delta n_{c d}^{\text{C} \rightarrow \text{V} \rightarrow \text{D}}   + \mathrm{P}_{\text{C} \rightarrow \text{V} \rightarrow \text{C}}^{i \rightarrow j} \Delta n_{c d}^{\text{C} \rightarrow \text{V} \rightarrow \text{C}}                                                           \\
		& +\mathrm{P}_{\text{D} \rightarrow \text{V} \rightarrow \text{D}}^{i \rightarrow j} \Delta n_{c d}^{\text{D} \rightarrow \text{V} \rightarrow \text{D}}+\mathrm{P}_{\text{D} \rightarrow \text{V} \rightarrow \text{C}}^{i \rightarrow j} \Delta n_{c d}^{\text{D} \rightarrow \text{V} \rightarrow \text{C}}),\nonumber
	\end{aligned}
\end{equation}
\begin{equation}
	\begin{aligned}
		\dot{p}_{c v}=
		& \frac{1}{N k} \sum_{n_{c}^{o}, n_{d}^{o} ; n_{c}^{m}, n_{d}^{m}}(\mathrm{P}_{\text{C} \rightarrow \text{D}}^{i \rightarrow i} \Delta n_{c v}^{\text{C} \rightarrow \text{D}}+\mathrm{P}_{\text{D} \rightarrow \text{C}}^{i \rightarrow i} \Delta n_{c v}^{\text{D} \rightarrow \text{C}}          \\
		& +\mathrm{P}_{\text{C} \rightarrow \text{V} \rightarrow \text{D}}^{i \rightarrow j} \Delta n_{c v}^{\text{C} \rightarrow \text{V} \rightarrow \text{D}}+        \mathrm{P}_{\text{C} \rightarrow \text{V} \rightarrow \text{C}}^{i \rightarrow j} \Delta n_{c v}^{\text{C} \rightarrow \text{V} \rightarrow \text{C}}                                                \\
		& +\mathrm{P}_{\text{D} \rightarrow \text{V} \rightarrow \text{D}}^{i \rightarrow j} \Delta n_{c v}^{\text{D} \rightarrow \text{V} \rightarrow \text{D}}+\mathrm{P}_{\text{D} \rightarrow \text{V} \rightarrow \text{C}}^{i \rightarrow j} \Delta n_{c v}^{\text{D} \rightarrow \text{V} \rightarrow \text{C}}),  \nonumber
	\end{aligned}
\end{equation}

\begin{equation}
	\begin{aligned}
		\dot{p}_{d d}=
		 & \frac{2}{N k} \sum_{n_{c}^{o}, n_{d}^{o} ; n_{c}^{m}, n_{d}^{m}}(\mathrm{P}_{\text{C} \rightarrow \text{D}}^{i \rightarrow i} \Delta n_{d d}^{\text{C} \rightarrow \text{D}}+\mathrm{P}_{\text{D} \rightarrow \text{C}}^{i \rightarrow i} \Delta n_{d d}^{\text{D} \rightarrow \text{C}}                                                      \\
		 & +\mathrm{P}_{\text{C} \rightarrow \text{V} \rightarrow \text{D}}^{i \rightarrow j} \Delta n_{d d}^{\text{C} \rightarrow \text{V} \rightarrow \text{D}}+    \mathrm{P}_{\text{C} \rightarrow \text{V} \rightarrow \text{C}}^{i \rightarrow j} \Delta n_{d d}^{\text{C} \rightarrow \text{V} \rightarrow \text{C}}        \\
		 & +\mathrm{P}_{\text{D} \rightarrow \text{V} \rightarrow \text{D}}^{i \rightarrow j} \Delta n_{d d}^{\text{D} \rightarrow \text{V} \rightarrow \text{D}}+\mathrm{P}_{\text{D} \rightarrow \text{V} \rightarrow \text{C}}^{i \rightarrow j} \Delta n_{d d}^{\text{D} \rightarrow \text{V} \rightarrow \text{C}}),\nonumber
	\end{aligned}
\end{equation}
where $\Delta n_{c c}^{\text{C} \rightarrow \text{D}}$ represents the variation of the number of $\text{CC}$ pairs caused by a cooperator staying put and imitating the strategy from a defector.
Analogously, $\Delta n_{c c}^{\text{C} \rightarrow \text{V} \rightarrow \text{C}}$ ($\Delta n_{c c}^{\text{C} \rightarrow \text{V} \rightarrow \text{D}}$) corresponds to the variation of the number of $\text{CC}$ pairs
caused by a cooperator migrating to an empty site and imitating strategy from a cooperator (defector).
And the variations for the number of $\text{CD}$, $\text{CV}$ and $\text{DD}$ pairs can be similarly obtained.
With initial pair probabilities $p_{c c}=p_{c}^{2}, p_{c d}=p_{c} p_{d}, p_{c v}=p_{c} p_{v}, p_{d d}=p_{d}^{2}$, the stable frequency of cooperators can be solved through numerical iterations.
And the frequency of cooperators $p_c$ at the system equilibrium can be obtained by the summation of equilibrium pair probabilities $p_{c c}$, $p_{c d}$ and $p_{c v}$.

\subsection{Theoretical analysis for evolutionary games with misinformation}
The evolutionary process with misinformation takes $p_{cc}^\text{t}$, $p_{cd}^\text{t}$, $p_{cv}^\text{t}$, $p_{dd}^\text{t}$, $p_{cc}^\text{m}$, $p_{cd}^\text{m}$, $p_{cv}^\text{m}$, $p_{dd}^\text{m}$, $p_{dv}^\text{m}$ to describe the collective interaction, which represent the pair configuration probabilities of $\text{CC}$, $\text{CD}$, $\text{CV}$ and $\text{DD}$ pairs in true and misinformation areas respectively.
Thus, $\mathrm{P}_{\text{D}^x \rightarrow \text{C}^x}^{i \rightarrow i}$, $\mathrm{P}_{\text{D}^x \rightarrow \text{V}^x \rightarrow \text{C}^x}^{i \rightarrow j}$, $\mathrm{P}_{\text{D}^x \rightarrow \text{V}^x \rightarrow \text{D}^x}^{i \rightarrow j}$ are calculated subsequently, where $x\in \{\text{t},\text{m}\}$, describing the migration and strategy change in area with true and misinformation.
The expected change rate of pair probabilities of $\text{CC}$, $\text{CD}$, $\text{CV}$ and $\text{DD}$ pairs are calculated analogously, and
the stable values for pair probabilities are obtained through numerical iterations.

\bibliography{bib_migration}

\end{document}